# ALMA Newsletter
February 2012

## ALMA in-depth

### How ALMA is calibrated:
### I. Antenna-based pointing, focus and amplitude calibration
by T. van Kempen, S. Corder, R. Lucas and R. Mauersberger.

There are several good reasons to frequently calibrate ALMA and its data: First, good calibration is necessary to obtain good quality images and reliable intensities and coordinates of sources in the sky. Second, it helps to save observing time and makes sure the observations are as efficient as possible. And finally, frequent calibrations help to monitor the system as a whole and to detect any imperfections or failures. Some of the calibrations are explicitly requested by the PI during the observation of a scheduling block. These include the calibration of the amplitude, the focus, the pointing relative to a source in a certain region of the sky, and the phase of the different antennas in an array. Other types of calibrations are constantly being performed during an observation without the need to specify those in the instructions for a scheduling block. Examples are the correction of phase drifts in the system by line length correctors, the determination of atmospheric phase variations via the Water Vapor Radiometers (WVR), and finally the calibration of the pointing behavior of the antennas via a metrology system. Finally, there are calibrations of the whole system that are performed when an antenna is integrated into the array after it has been moved from one pad to another and also on a regular basis (at present every week) by ALMA staff in order to monitor the system. The latter calibrations include the determination of a global pointing model for each antenna and each band, the determination of the exact position of each antenna in the array (the baseline determination), a check of the focus for each antenna and band, as well as a check of the receiver temperatures.

Here we will discuss which calibrations are needed, how to perform them and how the calibration affect ALMA observations. This first part concentrates on aspects related to a single antenna, namely antenna pointing (including the antenna metrology), focusing, and the amplitude

# ALMA Newsletter
February 2012

## ALMA in-depth

calibration, we also mention the bandpass calibration for single dish measurements. In a second part, we will explain those parts of the calibration that relate to the interferometer as a whole, and have to do with the phase stability of the system, namely the phase calibration, a more detailed description of the bandpass calibration and the baseline calibration. More exhaustive discussions can be found e.g. in Baars, (2007), Taylor, Carilly & Perley (1999), Thompson, Moran & Swenson (2001) and Wilson, Rohlfs & Hüttemeister (2008).

### ANTENNA POINTING

In order to maximize the signal received from a source and to minimize the calibration uncertainties, we have to be able to accurately track its position in the sky. Unlike optical telescopes, we have no guide stars in the field of view, and often the source we want to observe is only detectable after a long integration time. At the highest frequencies ALMA will be able to observe, namely almost 1 THz, the 12 m antennas of the main array of ALMA each have a half power beamwidth of 7". It is obvious that the blind pointing of the antennas should be much more accurate than such a beam size in order to find a source at such a high frequency. In fact, the specifications of the ALMA antennas require a "blind pointing" accuracy of 2" all over the sky. This specification is extremely challenging: after all, the antennas are not protected by a dome, and experience high wind and large temperature gradients by ambient temperature that can rapidly change and, during the day, from sunlight illuminating the antenna.

Each of the antenna vendors from the North American, European and East Asian Regions applied their own solutions to ensure such a high pointing accuracy. This includes a careful design, and modeling of the antennas' behavior in response to temperature, gravitation and wind, and also metrology devices in the antenna such as sensors that measure the position of the antenna's outer structure with respect to a reference frame inside an antenna. There are also thermal sensors and inclinometers which help to correct for wind gusts and thermal deformations. Still, this is not sufficient to guarantee that celestial sources can be pointed at with sufficient accuracy.

When an antenna is assembled at the Operation Support Facility (OSF) its mechanical behavior is characterized by observing a large number of celestial sources distributed all over the sky using the radio receivers in single dish, and later in interferometric mode. For each of the sources, the telescope is pointed first toward the position where the telescope control system supposes the source to be and then at half beam offsets in either direction in azimuth and elevation. During this process, the signal from the radio source is recorded, and from a Gaussian fit to the signal one can determine how much the azimuth and elevation encoder readings of the telescope deviate from the predicted azimuth and elevation of the source in the sky (taking the atmospheric refraction properly into account). We can then relate the azimuth and elevation

# ALMA Newsletter
February 2012

## ALMA in-depth

offsets Δaz(az, el), Δel(az, el), measured for hundreds of radio sources at different values of azimuth and elevation, to a model of the misalignments and imperfections any telescope has. This includes among others, the gravitational bending of the telescope as a function of elevation, the index errors in azimuth and elevation, the small tilt of the azimuth axis, and the fact that the neither the azimuth axis nor the line of sight are exactly perpendicular to the elevation axis.

This pointing model is then implemented into the control system of telescope in such a way that the antennas can now "blindly" find any source in the sky with an accuracy of 2" or better. Such a pointing model should be valid for several weeks before an upgrade of the pointing constants becomes necessary. A relatively short pointing run with 30-50 sources, spread evenly between 20 and 85 degrees elevation, is performed on a weekly basis to monitor the quality of the pointing model.

The antenna pointing model is sufficient to find any source in the sky, even at the highest frequencies. It is however not sufficiently well determined and stable to allow centering a source in the beam within 1/10 of the smallest beam width in order to optimize the calibration and sensitivity. Therefore, each scientific observation (Scheduling Block) is preceded by an offset pointing, i.e. a pointing observation on a strong point source close with well determined coordinates to the scientific target. Such an offset pointing allows to apply the small correction needed to track any source in the sky within the 0.6" specified in the ALMA requirements. This offset pointing can be performed at band 3 or 6 even if the actual science observations are performed. E.g. in band 9, where if is difficult to find suitable pointing sources.

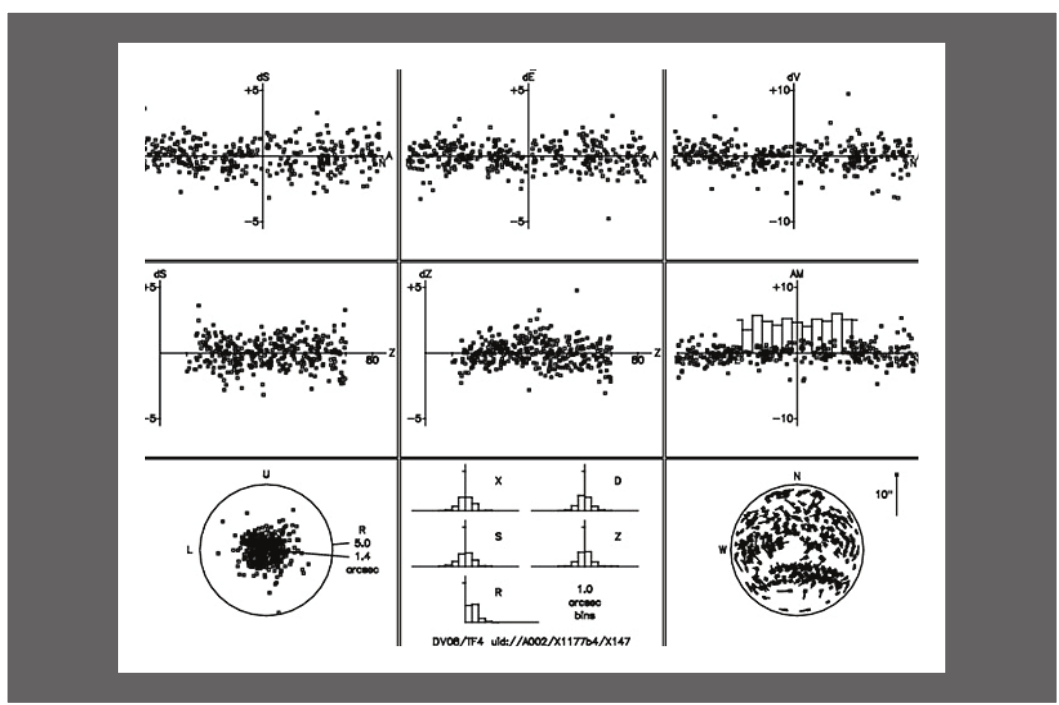

*After many celestial sources have been POINT is used to relate the offsets to about 15 pointing constants which determine the pointing model of an antenna. This pointing model predicts the imperfections in the aligment of an antenna such than any source in the sky can be pointed at with an rms precision of less than 2"*

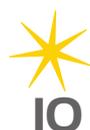

# ALMA Newsletter
February 2012

## ALMA in-depth

### ANTENNA FOCUS

The ALMA dishes are deliberately constructed to change their shape as a function of elevation in such a way that the antenna always has the form of a paraboloid, however with varying focal length. This elegant principle of homologous deformation allows us to build large telescopes with a lighter structure, but it requires us to refocus the telescope as a function of elevation. This is taken care of by the part of the antenna control system that is in charge of the subreflector movement. In addition, the axial and lateral movement of the subreflector depend on the temperature of the antenna structure. After an antenna is assembled, ALMA astronomers determine the gravitational and temperature dependence of the axial and lateral focus by observations of strong radio sources at different elevations and temperatures. In a focus observation, we point the antenna toward such a radio source (e.g. a quasar or a planet) and then move the subreflector slightly inwards and outwards (and also toward lateral offsets) of the nominal focus position. From a Gaussian fit to the recorded intensities, we can predict the optimum focus position. As a result, we have a pretty accurate focus model for each antenna, which, however, does not predict effects of unequal illumination by the sun or large temperature gradients within the telescope structure. These residual affects can be corrected for during scheduling block by including a focus determination on a strong radio source and applying a small correction on top of our focus model.

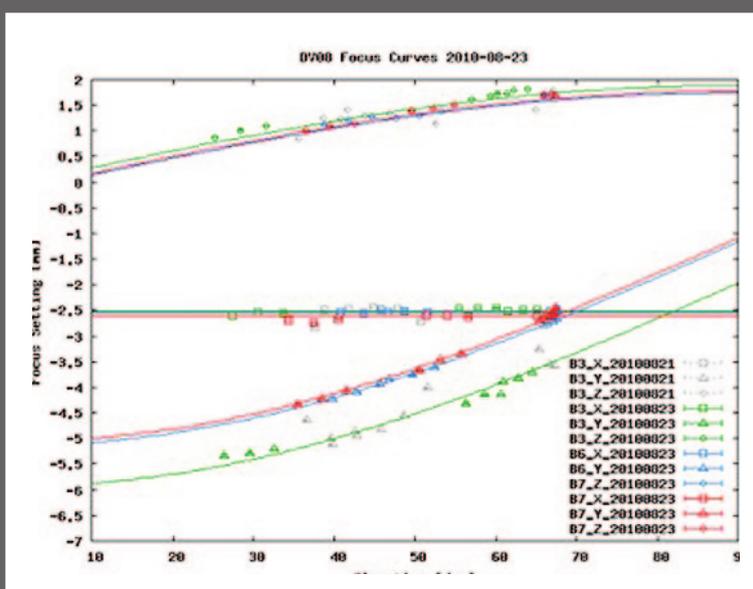

*This is a compilation of optimum focus settings (in three different focus directions) for one antenna at various elevations. The results of these measurements are used by the ALMA control software to automatically refocus the antenna during observations.*

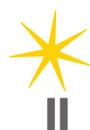

# ALMA Newsletter
February 2012

## ALMA in-depth

### AMPLITUDE CALIBRATION

ALMA is designed to yield data with an unprecedented flux density calibration uncertainty of 1% (3% at the highest frequencies). Even if we have managed to point up our antennas and focus them following the steps described above, this is an extremely challenging task. The reason is that we have to include a plethora of effects that threaten to degrade the calibration of the radiation of celestial source on its way through the atmosphere, the antenna, and the receivers into the detecting system. Closely related to the calibration of ALMA data is the monitoring of the noise produced by the receivers and the system, including the atmosphere, in order to track the performance of the equipment and to decide on observing strategies.

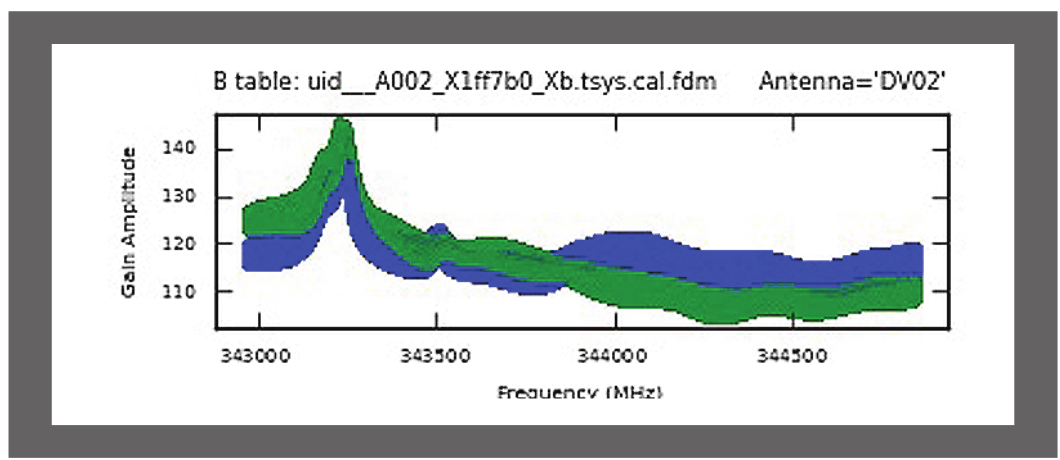

*The Figure shows a Tsys measurement on DV02 around 344 GHz. At 343.3 GHz, there is a known mesospheric line in the atmosphere. As such the system temperatures are about 20 to 30 K higher at these frequencies due to the lower transmission. Astronomical observations are consequently known to be less sensitive around 343.3 GHz.*

The basis of the ALMA calibration is the Alma Calibration Device (ACD). The ACD consists of robotic arms that can place two microwave absorbers, the ambient and the hot load, at well defined temperatures in front of the receiver feed horns of any of the bands. These loads have to work in the entire frequency range of 30 to 950 GHz foreseen for ALMA. The ambient load is at 20º C and the hot load at 70º C. The response of the system V is proportional to the temperature of the ambient or hot load plus the equivalent noise temperature of the receiver: $V_1=g*(T_{amb}+T_{rec})$ and $V_2=g*(T_{hot}+T_{rec})$. Thus a set of measurements of the response to a measurement toward an ambient load and toward a hot load can be used to determine at the same time the "receiver gain" g, which is the basis of the intensity scale for all calibrations of the ALMA system and ALMA data, and the receiver temperature, $T_{rec}$. The receiver temperature is a measure of the sensitivity of a receiver and it is important to monitor that it always complies with the specifications (see the article in ALMA Newsletter 6 http://www.almaobservatory.org/outreach/newsletter/211-newsletter-no6#alma_in_depth) over the whole observable frequency band.

The receiver is not the only part of the system that adds noise to the observations. The antenna and the atmosphere also contribute to the noise budget. The contribution of the atmosphere depends on the precipitable water vapor (PWV) toward the zenith, the elevation, and most importantly, the frequency. In particular, observations at the highest ALMA frequencies are only

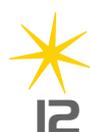

# ALMA Newsletter
**February 2012**

## ALMA in-depth

possible if the PWV content is small. Knowing the receiver temperature $T_{rec}$ and the gain g, we just have to point the antenna toward the sky to obtain the effective noise temperature of the atmosphere in the observing direction: $V_3 = g*(T_{sky}+T_{rec})$. In fact, the ALMA calibration software also accounts for the fact that a small percentage of the telescope beam is directed not toward the sky but toward the ground, and also for the contribution of the cosmic background radiation. It is important to know the noise temperature of the sky since, together with an atmospheric model, it can be related to the optical depth, $\tau$, of the atmosphere in the observing direction and at the observing frequency. The optical depth of the atmosphere not only determines the contribution of the sky to our noise budget but also the atmospheric absorption of the incoming radiation from a celestial source. The above mentioned calibrations also serve to calculate the system temperature during an observation. Since the system temperature contains a correction for the atmospheric absorption it can be related to a black body outside the atmosphere, and can therefore be used in the radiometer formula to estimate the integration time necessary to obtain a certain signal to noise ratio at a given frequency and elevation.

During a scientific observation, sky and receiver temperatures are measured every 10 to 20 minutes (depending on the observing band and weather conditions), and whenever one observes in a different direction of the sky (in practice >15º away). Results are stored together with the observed data and are used to apply corresponding corrections due to the atmospheric absorption, as one of the first steps in the data reduction process.

For a perfect antenna, this would already be sufficient to exactly calibrate all of our measurements. Since, however, the ALMA antennas do not perfectly couple to the sky, and since there might be still a residual mis-pointing and defocus, a thorough calibration requires sources in the sky for which we know the flux at any given frequency. Among the best such calibrators are solar system objects especially those for which we know the size and the temperature. In particular for Neptune, Uranus and Mars there is a good understanding of their fluxes as a function of frequency and time. Minor planets and the moons of the giant planets can also be used in particular Ganymede and Callisto around Jupiter, and Titan around Saturn.

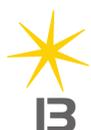

# ALMA Newsletter
February 2012

## ALMA in-depth

ALMA is designed to observe regularly at baselines longer than 1 km and as long as 16 km apart. At such baselines most solar system objects, including asteroids and moons, are resolved out at any frequency, making amplitude calibration difficult. One can use models of the coupling of the synthesized beam of the array to an extended sources, but this is dependent on the configuration of the array relative to the location of such an object in the sky at the time of the observations and may not always give the best results. Another obvious problem is that solar system object are unevenly distributed over the sky.

This is why ALMA uses quasars as secondary calibrators: they are point like and distributed all over the sky. A disadvantage of quasars is, however, that their fluxes may vary with time and that the frequency distribution of the flux, the spectral energy distribution, is not known a priori. Almost all known quasars spend significant amounts of time in a flaring state with brightnesses that can vary day by day. This is why ALMA is carrying out a monitoring program of 40 of the brightest visible quasars distributed all over the sky. The fluxes of the calibrators are calibrated against the fluxes of solar system objects, which serve as ALMA's primary calibrators. The cross calibration is performed using telescope pairs that are at a short distance of each other in order to capture all the flux from the primary calibrators. During the scientific observations, however, these quasars can be used at all baselines since they are compact.

The monitoring program is also not limited to a single frequency, but these quasars are observed at low frequency (currently 92 GHz) and high frequency (352 GHz) to determine the frequency dependency of the flux. With observations at multiple wavelengths, it is possible to estimate fluxes of these quasars at any frequency.

### BACKGROUND REMOVAL FOR SINGLE DISH MEASUREMENTS

ALMA will also be used for single dish measurements, in order to map structures that are "resolved out" by the interferometer. For spectral line measurements, the correlator will then be used in the autocorrelation mode with bandwidths of up to several GHz. The signal from a celestial source is typically hundreds or thousands time weaker that the contribution of the atmosphere. Since the atmospheric contribution to the signal varies with frequency (in fact there are many atmospheric absorption lines) and time, one has to frequently measure a blank portion of the sky in order to determine the contribution of the atmosphere to the observed signal. In practice one either observes a reference position close to the source observed every minute or so, or one uses emission free regions of a map when performing an on-the-fly map of a source.

# ALMA Newsletter
February 2012

## ALMA in-depth

### WATER VAPOUR RADIOMETRY

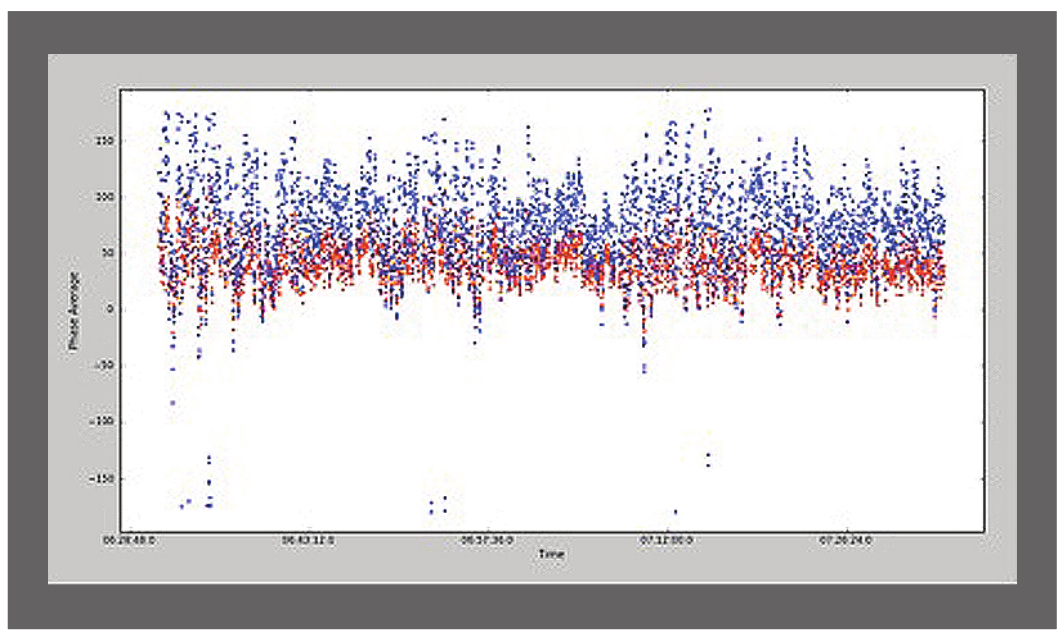

*Comparison between the phase stability of a quasar before correction of the WVR is applied (blue) and after (red). As can be seen the variation in phase is significantly reduced.*

ALMA's 12-meter antennas are equipped with water vapor radiometers (WVR; for a detailed description, see Newsletter http://www.almaobservatory.org/outreach/newsletter/211-newsletter-no6#focus_on). These are receivers which, in parallel to the receivers used for astronomy, look towards the sky and record the atmospheric emission at 183 GHz and neighboring frequencies. At this frequency, there is a strong atmospheric emission line of a rotational transition of water vapor. By measuring the intensity of that line emission in its central frequency and its line wings, one can calculate the amount of water vapor in the line of sight and the delay this water vapor causes to an incoming signal. This delay is used to correct for the phase fluctuations caused by atmospheric water vapor. These predicted phase corrections are continously recorded and can be used at a later stage of data reduction to reduce the phase variations between different antennas. This makes interferometric observations possible even for large baselines, high frequencies and less optimum weather conditions. The WVRs are also used in order to decide whether certain projects that require specially low water vapor in the atmosphere can be observed. The astronomers and telescope operators can see a display on the console showing the evolution of the water vapour in the atmosphere as a fuction of time.

# ALMA Newsletter
**February 2012**

## ALMA in-depth